\def\bk{{\bf k}}
\def\bkp{{\bf k'}}
\def\br{{\bf r}}
\def\bR{{\bf R}}
\def\bd{{\bf d}}
\def\w{\omega}
\def\wk{\omega_k}
\def\w0{\omega_0}
\def\ekj{\hat{e}_{\bk j}}
\def\ekjp{\hat{e}_{\bk 'j'}}

\def\ekj{\hat{e}_{\bk j}}

\documentclass[preprint,aps,showpacs]{revtex4}

\begin{document}

\title{Spatial correlations of vacuum fluctuations and the Casimir-Polder potential}

\author{R. Passante\mbox{${\ }^{*}$}, F. Persico\mbox{${\ }^{**}$}, L. Rizzuto\mbox{${\ }^{**}$}}
\affiliation{
\mbox{${\ }^{*}$} Istituto di Biofisica - Sezione di Palermo, 
Consiglio Nazionale delle Ricerche, Via Ugo La Malfa 153, I-90146 Palermo, Italy \\
\mbox{${\ }^{**}$} INFM and Dipartimento di Scienze Fisiche ed Astronomiche, 
Universit\'{a} degli Studi di Palermo, Via Archirafi 36, I-90123 Palermo, Italy}

\email{passante@iaif.pa.cnr.it}

\begin{abstract}
We calculate the Casimir-Polder intermolecular potential using an effective
Hamiltonian recently introduced. We show that the potential can be expressed
in terms of the dynamical polarizabilities of the two atoms and the equal-time
spatial correlation of the electric field in the vacuum state. This gives support
to an interesting physical model recently proposed in the literature, where
the potential is obtained from the classical interaction between the instantaneous
atomic dipoles induced and correlated by the vacuum fluctuations. Also, the
results obtained suggest a more general validity  of this intuitive model, for example
when external boundaries or thermal fields are present.
\end{abstract}


\maketitle

Casimir-Polder forces are long-range intermolecular forces between neutral atoms 
or molecules arising from their common interaction with the electromagnetic 
radiation field \cite{CP48,Milonni94}. They have also been measured with
good precision \cite{SBCSH93}. Casimir-Polder forces are among the few unambiguous
manifestations of the quantum nature of the electromagnetic field 
\cite{Spruch96,BAM00}, and they are
also the microscopic basis of the Casimir effect \cite{Milonni94,BMM99,GLR03,CGVC03}.
Thus it is very important to consider physical models which allow to understand
the origin of these forces,
and many efforts have been devoted to this subject 
(see \cite{CPP95} and references therein).
In the models proposed, the origin of the Casimir-Polder force is traced back to
the vacuum fluctuations or to the radiation reaction field, 
or to a combination of them.  

An interesting heuristic model has been proposed by Power and Thirunamachandran, which involves
the spatial correlations of vacuum fluctuations \cite{PT93}. Their picture, which
conceptually derives from the quantum theory of London dispersion forces, traces
the origin of the Casimir-Polder potential back to the instantaneous dipole moments 
of the two atoms which are induced by the vacuum
fluctuations. The induced dipoles are correlated because vacuum fluctuations
have spatial correlations \cite{CDG89}, and this yields a nonvanishig average dipole-dipole 
interaction, although the average value of the induced dipoles vanishes. 
Using this picture, Power and Thirunamachandran in \cite{PT93} were able to obtain
the complete Casimir-Polder potential, both in the socalled near and 
far zones. This physically transparent approach to the Casimir-Polder potential
has also been generalized to the three-body component of the potential in
the case of three ground state atoms in vacuo \cite{CP97}.

In this paper, we show that the above mentioned approach to the Casimir-Polder potential
in terms of the spatial correlations
of vacuum correlations, originally introduced as an intuitive physical
model, can be derived rigorously from the Hamiltonian of the two atoms
interacting with the radiation field.
This gives further support to the validity of this model, and also indicates
that this validity can be further extended 
for example to the case of the Casimir-Polder potential
at finite temperature or in the presence of boundary conditions. 
Casimir-Polder interactions at finite temperature or with external boundaries 
have recently received great attention in the
literature \cite{WDN99,Barton01,KMM00}.

In the multipolar coupling scheme and within the dipole approximation, the
Hamiltonian describing two atoms A and B interacting with the radiation
field is

\begin{equation}
H = \sum_{\bk j} \hbar \wk a_{\bk j}^\dagger a_{\bk j} + H_{atom}^A +H_{atom}^B
-\mbox{\boldmath $\mu_A$} \cdot \bd (\br_A)
-\mbox{\boldmath $\mu_B$} \cdot \bd (\br_B)
\label{eq:1}
\end{equation}
where $\br_A$, $\br_B$ are the positions of the two atoms,
$\mbox{\boldmath $\mu$}$ indicates the atomic
dipole moment  and
$\bd (\br ) = \sum_{\bk j} \bd_{\bk j}(\br )$ is the transverse displacement field,
which, in this coupling scheme, is the momentum conjugate
to the vector potential \cite{CPP95}.

We are interested in the Casimir-Polder potential for a pair ground-state atoms.
In \cite{PPT98} it has been shown that this kind of calculation can be
considerably simplified using an effective Hamiltonian that
can be obtained from the Hamiltonian (\ref{eq:1}).  This effective Hamiltonian, which is
equivalent to the original one, is quadratic in the field operators; the effective interaction
term is
\begin{equation}
H_{int} = H_{int}^A + H_{int}^B = 
-\frac 12 \sum_{\bk j} \alpha_A (k) \bd_{\bk j}(\br_A) \cdot \bd (\br_A)
-\frac 12 \sum_{\bk j} \alpha_B (k) \bd_{\bk j}(\br_B) \cdot \bd (\br_B)
\label{eq:2}
\end{equation}
$\alpha (k)$ is the atomic dynamic polarizability
\begin{equation}
\alpha (k) = \frac 2{3 \hbar c} \sum_m \frac {k_{m0} \mid \mbox{\boldmath $\mu^{m0}$}
\mid^2}{k_{m0}^2 - k^2}
\label{eq:3}
\end{equation}
where $\hbar ck_{m0} = E_m - E_0$ is the transition energy from the atomic state $m$
to the ground state $0$ and $\mbox{\boldmath $\mu^{m0}$}$
are the matrix elements of the atomic dipole moment operator.

We can calculate the Casimir-Polder interatomic potential in two steps: 
we first obtain the dressed ground
state of one atom (A), as if the other atom were absent, and then we evaluate the interaction
between atom B and the virtual photon cloud dressing atom A. 
This second step is
equivalent to calculating the average value of the 
effective interaction Hamiltonian $H_{int}^B$ pertaining to atom B 
on the dressed ground state of A. A similar approach
has already been used for the special case of the Casimir-Polder potential in the socalled
{\it far zone} \cite{PP87}.

At the lowest order in the atomic polarizability, the dressed ground state of atom A can be
obtained by a straightforward perturbation theory

\begin{equation}
\mid \tilde{g}_A \rangle = \mid g_A 0_k \rangle 
- \frac \pi V\sum_{\bk j \bkp j'} \alpha_A(k)
\ekj \cdot \ekjp \frac {(kk')^{1/2}}{k+k'} e^{-i(\bk+\bkp) \cdot \br_A} \mid g_A \bk j \bkp j' \rangle
\label{eq:4}
\end{equation}
where $g_A$ indicates the bare ground state of atom A and $\bk j$ are photon states.
In order to calculate the expectation value of the effective Hamiltonian (\ref{eq:2}) 
describing the interaction of atom B with
the radiation field, we first evaluate the following quantity

\begin{eqnarray}
&\ & \langle \tilde{g}_A \mid \bd_{\bk j}(\br_B) \cdot \bd (\br_B) \mid \tilde{g}_A \rangle
= \frac {2\pi^2 \hbar c}{V^2} \sum_{\bkp j'} \frac {kk'}{k+k'} 
\left( \alpha_A (k) +  \alpha_A (k') \right) 
\nonumber \\
&\times& \left( \ekj \cdot \ekjp \right)^2 e^{i(\bk +\bkp ) \cdot (\br_B - \br_A)} + c.c.
\label{eq:5}
\end{eqnarray}

We now use this expression to obtain the interaction energy between the two atoms
A and B, given by

\begin{equation}
E_{int} = -\frac 12 \sum_{\bk j} \alpha_B (k)
\langle \tilde{g}_A \mid \bd_{\bk j}(\br_B) \cdot \bd (\br_B) \mid \tilde{g}_A \rangle
\label{eq:6}
\end{equation}
After some algebra, we obtain

\begin{eqnarray}
E_{int} &=& -\frac{2\pi^2 \hbar c}V \sum_{\bk j} k \left( \ekj \right)_m
\left( \ekj \right)_\ell e^{i\bk \cdot \bR} 
\left( \frac 1V
\alpha_A(k) \sum_{\bkp j'} \frac {k'^2}{k'^2-k^2} \left( \ekjp \right)_m
\left( \ekjp \right)_\ell e^{i\bkp \cdot \bR} \alpha_B(k')
\right. \nonumber \\
&+& \left. \frac 1V
\sum_{\bkp j'} \frac {k'^2}{k'^2-k^2} \left( \ekjp \right)_m
\left( \ekjp \right)_\ell e^{i\bkp \cdot \bR} \alpha_A(k') \alpha_B(k')
\right) + c.c.
\label{eq:7}
\end{eqnarray}
where ${\bf R} = \br_B - \br_A$ is the distance between the two atoms.
After evaluation of polarization sums and angular integrations, the
term inside the main brackets in (\ref{eq:7}) becomes

\begin{equation}
\frac 1{2\pi^2} F_{\ell m}^R \frac 1R \left(
\alpha_A(k) \int_0^\infty \! dk' \frac {k'}{k'^2-k^2} \sin k'R \: \alpha_B(k')
+ \int_0^\infty \! dk' \frac {k'}{k'^2-k^2} \sin k'R \: \alpha_A(k') \alpha_B(k')
\right)
\label{eq:8}
\end{equation}
where 

\begin{equation}
F_{\ell m}^R = \left( \nabla^2 \delta_{\ell m} - \nabla_\ell \nabla_m \right)^R
\label{eq:8a}
\end{equation} 
is a differential operator acting on the coordinate ${\bf R}$.

Using $\alpha (k) = \alpha (-k)$ and the analytical properties of the
dynamical polarizability, the following integrals are easily evaluated

\begin{eqnarray}
\int_0^\infty \! dk' \frac {k'}{k'^2-k^2} \sin k'R \:  \alpha_B(k')
&=& \frac \pi 2 \alpha_B(k) \cos (kR) 
\label{eq:9} \\
\int_0^\infty \! dk' \frac {k'}{k'^2-k^2} \sin k'R \: \alpha_A(k') \alpha_B(k')
&=& \frac \pi 2 \alpha_A(k) \alpha_B(k)  \cos (kR) 
\label{eq:10}
\end{eqnarray}
and eq. (\ref{eq:8}) becomes

\begin{equation}
\frac 1{2\pi} \alpha_A(k) \alpha_B(k) F_{\ell m}^R \frac {\cos kR}R
\label{eq:11}
\end{equation}

Substitution into eq. (\ref{eq:7}) yields the following expression for
the interaction energy between the two atoms

\begin{equation}
E_{int} = \frac {2\pi \hbar c}V \Re \left( \sum_{\bk j} \left( \ekj \right)_m
\left( \ekj \right)_\ell k e^{i\bk \cdot \bR} \alpha_A(k)
\alpha_B(k) \left( - F_{\ell m}^R \frac {\cos kR}R \right) \right)
\label{eq:12}
\end{equation}

Inspection of the form (\ref{eq:12}) of the Casimir-Polder potential shows
that it can be written in the following form

\begin{equation}
E_{int} = \Re \left( \sum_{\bk j} \left( 
\langle 0_k \mid \left( d_{\bk j}(\br_B) \right)_m \left( d_{\bk j}(\br_A) \right)_\ell
\mid 0_k \rangle
\alpha_A(k) \alpha_B(k) V_{\ell m}(k, \bR = \br_B - \br_A) \right) \right)
\label{eq:13}
\end{equation}
where

\begin{equation}
\langle 0_k \mid \left( d_{\bk j}(\br_B) \right)_m \left( d_{\bk j}(\br_A) \right)_\ell 
\mid 0_k \rangle
=  \frac {2\pi \hbar c}V \left( \ekj \right)_m \left( \ekj \right)_\ell k e^{i\bk \cdot \bR}
\label{eq:14}
\end{equation}
is the equal-time spatial correlation function of the electrical field
in the (bare) vacuum state and evaluated at the position of the two atoms.
The quantity

\begin{eqnarray}
&\ & V_{\ell m}(k, \bR ) = -F_{\ell m}^R \frac {\cos kR}R 
\nonumber \\
&=& k^3 \left(
\left( \delta_{\ell m} -\hat{R}_\ell \hat{R}_m \right)
\frac {\cos kR}{kR} - \left( \delta_{\ell m} -3\hat{R}_\ell \hat{R}_m \right)
\left( \frac {\sin kR}{k^2R^2} + \frac {\cos kR}{k^3R^3} \right) 
\right)
\label{eq:15}
\end{eqnarray}
is the classical electrostatic interaction energy between two dipoles oscillating at
frequency $ck$ \cite{MLP66}.

The form (\ref{eq:13}) of the Casimir-Polder potential, which we have derived
from the basic Hamiltonian of the atom-radiation interaction, 
coincides with the heuristic expression assumed on 
the basis of physical considerations by Power and
Thirunamachandran in \cite{PT93}. Therefore, our results
justify the validity of their physical model, which traces the origin of
the Casimir-Polder potential back to the classical electric interaction between
the two atomic dipoles, induced and correlated by the vacuum
fluctuations. Furthermore, our results, although explicitly derived 
for the case of two atoms in vacuo
at zero temperature, suggest that the model might be valid also in
more complicated situations, such as in the presence of boundary conditions
(conducting plates or dielectrics) or at nonzero temperature. In fact, we expect that
in these cases we should simply substitute the spatial correlation
of the electric field in the vacuum state with that calculated with the specific 
boundary conditions considered or in a thermal state of the field.

\acknowledgments

This work was supported by the European Commission under contract No. HPHA-CT-2001-40002
and in part by the bilateral Italian-Japanese project 15C1 on Quantum
Information and Computation of the Italian Ministry for Foreign Affairs. 
Partial support by Ministero dell'Universit\'{a} e della Ricerca Scientifica 
e Tecnologica and by Comitato Regionale di Ricerche Nucleari e di Struttura della Materia
is also acknowledged.

\end{document}